\documentclass[aps,prc,preprint,tightenlines,groupedaddress,nofootinbib,showpacs,preprintnumbers,amsmath,amssymb,superscriptaddress]{revtex4} 
\usepackage{graphicx}
\usepackage{dcolumn}
\usepackage{bm}
\usepackage{amsmath} 
 
\begin{document} 
 
\def\R{\right} \def\L{\left} \def\Sp{\quad} \def\Sp2{\qquad} 
 
\title{Strange-quark contribution to the ratio of neutral- to 
       charged-current cross sections in neutrino-nucleus scattering} 


\author{B. I. S. van der Ventel}
\email{bventel@sun.ac.za}
\affiliation{Department of Physics, Stellenbosch University,
Stellenbosch 7600, South Africa}

\author{J. Piekarewicz}
\email{jorgep@csit.fsu.edu}
\affiliation{Department of Physics, Florida State University, Tallahassee 
FL 32306, USA}
\date{\today} 
 
\begin{abstract} 
A formalism based on a relativistic plane wave impulse approximation
is developed to investigate the strange-quark content ($g_{A}^{s}$) of
the axial-vector form factor of the nucleon via neutrino-nucleus
scattering.  Nuclear structure effects are incorporated via an
accurately calibrated relativistic mean-field model. The ratio of
neutral- to charged-current cross sections is used to examine the
sensitivity of this observable to $g_{A}^{s}$. For values of the
incident neutrino energy in the range proposed by the FINeSSE
collaboration and by adopting a value of $g_{A}^{s}\!=\!-0.19$, 
a 30\% enhancement in the ratio is observed relative to the
$g_{A}^{s}\!=\!0$ result.
\end{abstract} 
 
 
\maketitle 

\section{Introduction} 
\label{section_introduction} 

In a desperate effort to salvage the conservation of energy and
angular momentum, Wolfgang Pauli postulated the existence of the
neutrino: a neutral and (almost) massless particle that he feared will
never be detected. Seventy five years later and with three neutrino
families firmly established, neutrino physics in the 21st century has
taken center stage in fields as diverse as cosmology, astro, nuclear,
and particle physics. With the commission of new neutrino
observatories and facilities for the study of solar, atmospheric, and
reactor neutrinos, conclusive evidence now exists in favor of neutrino
oscillations. Although neutrino-oscillation experiments have now
evolved from the discovery into the precision phase, important
questions remain unanswered~\cite{Kayser_NPB118_2003}. Depending on
these answers, a radical modification to the standard model of
particle physics may be required. At the very least, neutrino
oscillations already demand a mild extension to the standard model: 
in the standard model the {\it individual} lepton numbers must be
conserved.

The discovery of neutrino oscillations established two
incontrovertible facts: a) that neutrinos have mass and that these
masses can not all be equal and b) that the three known neutrinos
$\nu_{e}$, $\nu_{\mu}$, and $\nu_{\tau}$ are linear combinations of
three neutrino mass eigenstates (commonly referred as $\nu_{1}$,
$\nu_{2}$, and $\nu_{3}$). Note that due to the very small neutrino
masses, oscillation experiments are sensitive to only the {\it squared
mass difference} ($\Delta m^{2}$) of the mass eigenstates. Further,
the unitary matrix linking the flavor to the mass eigenstates has a
well-known counterpart in the quark sector: the CKM matrix. While
precision experiments have started to determine squared mass
differences and mixing angles, a strong, vibrant, and
interdisciplinary program is now being established to tackle the
myriad of remaining open questions~\cite{Kayser_NPB118_2003}.  Among
these are: What is the mass hierarchy? Is the neutrino a Dirac or
Majorana particle? Does the neutrino matrix contain a CP-violating
phase that may explain our matter-dominated Universe? Is there a need
for additional ``sterile'' neutrinos?

The apparent need for an additional sterile neutrino stems from two
older experiments. The first one in 1989 measured the total and
partial (into hadrons and charged leptons) widths of the $Z^{0}$ boson
at the large electron-positron (LEP) collider at CERN and extracted
the number of neutrinos flavors to be $N_{\nu}\!=\!3.00\pm0.08$; 
note that a more recent analysis reports
$N_{\nu}\!=\!2.984\pm0.008$~\cite{Eid04_PLB592}. The second
experiment was a 1995 neutrino-oscillation experiment at the Liquid
Scintillator Neutrino Detector (LSND) at the Los Alamos National
Laboratory~\cite{Athanassopoulos_PRL75_95}. This experiment reported
evidence of $\bar{\nu}_{\mu}\rightarrow\bar{\nu}_{e}$ oscillation, but
with a squared mass difference that is too large --- and thus
inconsistent --- with the two {\it independent} values extracted from
solar, atmospheric, and reactor experiments. Simply put, for three
neutrino flavors the algebraic relation $\Delta m_{21}^{2}\!+\!\Delta
m_{32}^{2}\!=\!\Delta m_{31}^{2}$ must be satisfied, yet $\Delta
m_{\rm sol}^{2}\!+\!\Delta m_{\rm atm}^{2} \!\neq\!\Delta m_{\rm
LSND}^{2}$. The most favorable scenario that accommodates three
independent $\Delta m^{2}$ values is the addition of a sterile
neutrino. Yet other possibilities exist: the LSND analysis may be
incorrect. It is the primary goal of the Booster Neutrino Experiment
(BooNE and its first phase MiniBooNE) at Fermilab to confirm the LSND
result.

While MiniBooNE's primary goal is to confirm the LSND result, this
unique facility is also ideal for the study of supernova neutrinos,
neutrino-nucleus scattering, and hadronic structure.  An ambitious
experimental program --- the Fine-grained Intense Neutrino Scattering
Scintillator Experiment (FINeSSE) --- aims to measure the
strange-quark contribution to the spin of the
nucleon~\cite{Potterveld_02,Bugel_03,Brice_2004}. FINeSSE is part of a
larger program started in the late 80's that attempts to answer a
fundamental nucleon-structure question: how do the non-valence
(``sea'') quarks --- particularly the strange quarks --- contribute to
the observed properties of the nucleon?  To date, the role of strange
quarks in the nucleon remains a contentious issue and one that remains
a subject of intense activity all over the world. In an attempt
to find a satisfactory answer to this fundamental question, a number of
reactions have been proposed. These include: (i) deep inelastic
scattering of neutrinos on protons
\cite{Bazarko_ZPhysC65_1995,Goncharov_PRD64_2001}, (ii) deep inelastic
scattering of polarized charged leptons~\cite{Adams_PRD56_1997}, (iii)
pseudoscalar meson scattering on a proton~\cite{DF90}, and (iv)
parity-violating electron scattering~\cite{McK89_PLB219,Beck89_PRD39}.
Part of the controversy arises because these reactions do not all
reach similar conclusions. For example, both reactions (i) and (ii)
suggest a non-zero strangeness contribution, in contrast to reaction
(iv) that indicates a strange quark contribution to the charge and
magnetic moment consistent with zero. Parity-violating electron
scattering in particular has received extensive experimental
attention as in the SAMPLE Collaboration at the MIT-Bates
accelerator~\cite{Mue97_PRL78}, the HAPPEX Collaboration at the
Jefferson Laboratory~\cite{Ani82_PRL82}, and the A4 Collaboration at
the MAMI facility in Mainz~\cite{Maas_EPJA17_2003}. However,
theoretical investigations have shown that large radiative
corrections and nuclear-structure effects impact negatively on the
extraction of strange-quark matrix elements
\cite{Musolf_PLB242_90,Musolf_PhysRep239_94,Horowitz_PRC47_93}.

Neutrino-induced reactions provide a viable alternative to
parity-violating electron scattering. While the latter is primarily
sensitive to the strange electric and magnetic form factors of the
nucleon, the former is particularly sensitive to the axial-vector form
factor of the proton --- through the combination $(\Delta u - \Delta d
- \Delta s + \Delta c - \Delta b + \Delta t)$. This sensitivity is the
result of the small weak-vector charge of the proton
($1-4\sin^{2}\theta_{\rm W}\!\simeq\!0.08$) and the suppression of the
weak anomalous magnetic moment at small $Q^{2}$. In the above
expression the heavy quark flavors ($c$, $b$, and $t$) can be
eliminated using a well-defined renormalization group
procedure~\cite{Bass_PRD66_2002,Bass_PRD68_2003}. Further, the
isovector combination $(\bar{u}\gamma_{\mu}\gamma_{5}u\!-\!
\bar{d}\gamma_{\mu}\gamma_{5}d)$ is constrained from neutron beta
decay. This leaves the (assumed isoscalar) strange-quark contribution
to the spin of the proton $\Delta s$ to be determined from the elastic
neutrino-proton reaction. Yet an absolute cross-section measurement of
this reaction is an experimental challenge due to difficulties in the
determination of the absolute neutrino flux. An attractive alternative
has been proposed by Garvey and
collaborators~\cite{Garvey_PLB289_92,Garvey_PRC48_93a} in which the
extraction of $\Delta s$ proceeds through a measurement of the {\it
ratio} of proton-to-neutron cross sections in neutral-current (NC)
neutrino-nucleon scattering. This ratio is defined by the following
expression:
\begin{equation}
\label{eq_1}
 R(p/n) =  \frac{\sigma(\nu p \rightarrow \nu p)} 
                {\sigma(\nu n \rightarrow \nu n)}\;.
\end{equation}
This ratio is very sensitive to the strange-quark contribution to the 
spin of the nucleon as $\Delta s$ 
[or $g_{A}^{s}\!\equiv\!G_{A}^{(s)}(Q^{2}\!=\!0)$] 
interferes with the isovector contribution ($G_{A}^{(3)}$) with one 
sign in the numerator and with the opposite sign in the denominator 
[see Eq.~\ref{eq_36}]. Unfortunately, $R(p/n)$ is difficult to measure 
with the desired accuracy due to experimental difficulties associated 
with neutron detection~\cite{Brice_2004}. It is for this reason that 
FINeSSE will focus initially on the neutral- to charged-current ratio 
(NC/CC):
\begin{equation}
\label{eq_2}
 R(NC/CC) = \frac{\sigma(\nu p \rightarrow \nu p)} 
                 {\sigma(\nu n \rightarrow \mu^{-} p)}\;.
\end{equation}
This ratio is ``simply'' determined from counting the number of events
with an outgoing proton and missing mass relative to those events with
an outgoing proton and a muon. Note that the CC reaction, being purely
isovector, is insensitive to $\Delta s$. As such, $ R(NC/CC)$ is
about a factor of two less sensitive to $\Delta s$ than $R(p/n)$.

From a theoretical perspective extracting $\Delta s$ from the ratio of
cross sections is also attractive. As a large number of the scattering
events at FINeSSE will be from nucleons bound to a Carbon nucleus, it
is important to understand nuclear-structure corrections.  This issue
came to light in experiment E374 at the Brookhaven National Laboratory
(BNL) where it was found that 80\% of the events involved neutrino
scattering off carbon atoms, while only 20\% were from free protons.
As nuclear-structure corrections in $R(NC/CC)$ appear to be
insensitive to final-state interactions between the outgoing proton
and the residual nucleus~\cite{Meucci_2004,Mar05_xxx0505008}, the
ratio $R(NC/CC)$ may be accurately computed using a much simpler
plane-wave formalism. Indeed, in Sec.~\ref{section_formalism} we will
show how the cross section ratio $R(NC/CC)$ in Carbon computed in a
plane-wave formalism may be expressed in a form that closely resembles
the ``Feynman-trace'' approach used to calculate the cross section
from free nucleons. We note in closing that the new generation of
neutrino experiments will require a thorough understanding of
neutrino-nucleus interactions since the detectors often contain
complex nuclei. Experimental and theoretical work related to neutrino
scattering from light and heavier nuclear targets may be found in
Refs.~\cite{Barish_PRD16_77,Athanassopoulos_PRC55_97,
Nakamura_PRC63_2001,Auerbach_PRC66_2002,Mintz_PRC40_1989,Kosmas_PRC53_1996,
Volpe_PRC65_2002,Maieron_PRC68_2003}.

Our paper has been organized as follows.  In
Sec.~\ref{section_formalism} we briefly review the formalism developed
in Ref.~\cite{VanderVentel_PRC69_2004} for the neutral-current case
and point out the main modifications required to make it applicable to
charged-current neutrino-nucleus scattering.  Our main results ---
with a focus on the sensitivity of $R(NC/CC)$ to $\Delta s$ --- are
presented in Sec.~\ref{section_results}.  Finally, we summarize the
main points of this work in Sec.~\ref{section_summary}.

\maketitle 
\section{Formalism} 
\label{section_formalism}

In this section the formalism for the description of charged-current
neutrino-nucleus scattering is presented. As the basic outline follows
closely the neutral-current formalism developed in
Ref.~\cite{VanderVentel_PRC69_2004}, we present a brief review that 
focusses on those modifications that arise from a finite muon mass.

\subsection{Cross section in terms of the leptonic and hadronic tensors}
\label{section_xsection_lepton_hadron}

\begin{figure}
\includegraphics[height=5cm,angle=0]{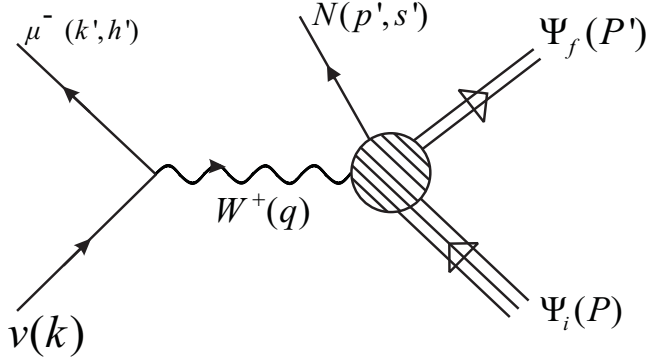}
 \caption{Lowest-order Feynman diagram for charged-current 
          neutrino-nucleus scattering.} 
\label{fig_1}
\end{figure}

The lowest-order Feynman diagram for the knockout of a bound proton
via the charged-current reaction $[\nu + X(Z,A) \rightarrow \mu^{-} +
p + X(Z\!-\!1,A\!-\!1)]$ is shown in Fig.~\ref{fig_1}. Here the
initial four-momentum of the (left-handed) neutrino is $k$ while the
four-momentum and helicity of the outgoing muon are $k'$ and $h'$,
respectively. The reaction proceeds via the exchange of a virtual
$W^{+}$ boson with four-momentum $q\!=\!(\omega,{\bf q})$. The
kinematical variables defining the hadronic arm are the four-momentum
of the target ($P$) and residual nucleus ($P'$). Finally, $p'$ and
$s'$ denote the four-momentum and spin component of the ejectile
proton. Energy-momentum conservation demands that:
\begin{equation}
  q=k-k'=p'+P'-P \;.
 \label{EMCons}
\end{equation}
The dynamical information for this reaction is contained in the 
transition matrix element given by
\begin{eqnarray}
\nonumber
 -i {\cal M}
 & = &
 \left\{ \overline{\mu} ({\bf k}', h') \,
  \left[ \displaystyle \frac{-ig}{2 \sqrt{2}} \, 
   \left( \gamma^{\mu} - \gamma^{\mu} \, \gamma^{5}
   \right) \,
  \right] \nu ({\bf k} \, )
 \right\} \, 
i {\cal D}_{\mu \nu} (q) \, 
\\
\label{eq_3}
 & &
 \left\{ \, \langle \, p',s' \, ; \; \Psi_{f} (P') \, \Big | \, 
 \displaystyle \frac{-ig}{2\sqrt{2}}\cos\theta_{C} \, 
  \hat{J}^{\nu} (q) \, \Big | \, 
 \Psi_{i} (P) \, \rangle \right\}.
\end{eqnarray}
In Eq. (\ref{eq_3}) the initial and final nuclear states are denoted by
$\Psi_{i}(P)$ and $\Psi_{f}(P')$, respectively. Furthermore, $g$ is the
weak coupling constant, $\theta_{C}$ is the Cabbibo angle 
($\cos\theta_{C}\!=\!0.974$), and $\hat{J}^{\nu} (q)$ is the weak 
nuclear current operator. As only low-momentum transfers 
($-q_{\mu}q^{\mu}\!\equiv\!Q^{2}\!\ll\!M_{W}^{2}$) will be considered, 
the following approximation is valid
\begin{equation}
\label{eq_4}
  {\cal D}_{\mu\nu}(q)=
  \frac{-g_{\mu\nu}+q_{\mu}q_{\nu}/M_{W}^{2}}
       {q^{2}-M_{W}^{2}}
  \longrightarrow  
  \frac{g_{\mu\nu}}{M_{W}^{2}}\;.
\end{equation}
Using this expression the transition matrix element may be written as
\begin{equation}
\label{eq_5}
 {\cal M} = 
  \frac{G_{F}}{\sqrt{2}}\cos\theta_{C} 
  \left[
   \overline{\mu}({\bf k}',h')\gamma_{\mu}(1-\gamma^{5})\nu({\bf k})
  \right]  
  \left[\langle p',s';\Psi_{f}(P')\Big| \hat{J}^{\mu}(q)\Big |
        \Psi_{i}(P)\rangle\right] \;,
\end{equation}
where the Fermi coupling constant $G_{F}$ has been introduced via
\begin{equation}
\label{eq_6}
 \frac{G_{F}}{\sqrt{2}}=\frac{g^{2}}{8 M_{W}^{2}}\;.
\end{equation}
In Eq.~(\ref{eq_5}) $\mu({\bf k}',h')$ is the Dirac spinor for the
outgoing muon expressed in the helicity representation. That is,
(suppressing ``prime'' indices for clarity),
\begin{equation}
 \mu({\bf k},h)=\displaystyle{
                \sqrt{\frac{E_{k}+m}{2E_{k}}}}
                  \begin{pmatrix}
                     \phi_{h}(\widehat{\bf k}) \\[0.25cm]
                     \displaystyle{
                      h \frac{k}{E_{k}+m}
                     \phi_{h}(\widehat{\bf k})}
		  \end{pmatrix}\;,
 \quad \Big(E_{k}=\sqrt{k^{2}+m^{2}}\;;\, k=|{\bf k}|\Big) \;,
 \label{eq_7}
\end{equation}
where $\phi_{h=\pm1}(\widehat{\bf k})$ are two-component Pauli 
spinors given by
\begin{equation}
  \phi_{h=+1}(\widehat{\bf k})=
    \begin{pmatrix}
     \cos(\theta/2) \\[0.25cm]
     \sin(\theta/2)\,e^{i\phi} 
  \end{pmatrix}\;,\quad
  \phi_{h=-1}(\widehat{\bf k})=
    \begin{pmatrix}
    -\sin(\theta/2)\,e^{-i\phi} \\[0.25cm]
     \cos(\theta/2)
  \end{pmatrix}\;.
 \label{eq_8}
\end{equation}
Here $m$ denotes the muon mass, and $\theta$ and $\phi$ are the polar 
and azimuthal angles of the muon momentum. The neutrino spinor
$\nu({\bf k})$ is directly obtained from the above expressions by
setting the fermion mass to $m\!=\!0$ and the helicity to $h\!=\!-1$. 
Note that left/right projection operators on plus/minus helicity states 
do not vanish in general due to the finite muon mass. That is,
 \begin{equation}
  {\cal P}_{L/R}\,\mu({\bf k},h\!=\!\pm1) \equiv
  \frac{1}{2}(1\mp\gamma^{5})\mu({\bf k},h\!=\!\pm1)=
  {\cal O}(m/k)\neq 0\;.
 \label{Projections}
\end{equation}
Further, we have adopted a non-covariant normalization for the
Dirac spinors of Eq.~(\ref{eq_7}),
\begin{equation}
\label{eq_9}
 \mu^{\dagger}({\bf k},h)\mu({\bf k}, h')=
 \bar{\mu}({\bf k},h)\gamma^{0}\mu({\bf k}, h')=
 \delta_{hh'} \;,
\end{equation}
a choice that is in accordance with the standard normalization of
the bound-state spinor~\cite{SW86} and that is given by
\begin{equation}
\label{eq_10}
 \int {\cal U}_{\alpha}^{\dagger}({\bf r})\,
      {\cal U}_{\alpha}({\bf r})\,d^{3}{\bf r}=1\;. 
\end{equation}
Following Ref.~\cite{VanderVentel_PRC69_2004} the differential cross 
section can now be written as
\begin{equation}
\label{eq_11}
 d \sigma = 
 \frac{G_{F}^{2}\cos^{2}\theta_{C}}
 {2(2 \pi)^{5}}d^{\,3}{\bf k'}d^{\,3}{\bf p'} 
 \delta (E_{k} + M_{A} - E_{k'} - E_{p'} - E_{P'}) 
 {\ell}_{\mu\nu} W^{\mu\nu} \;,
\end{equation}
where the leptonic tensor is given by
\begin{equation}
\label{eq_12}
 {\ell}_{\mu\nu} = {\rm Tr} 
  \left[\left(\gamma_{\mu} - \gamma_{\mu}\gamma_{5}\right)
        \Big(\nu({\bf k})\overline{\nu}({\bf k})\Big)
        \left(\gamma_{\nu} - \gamma_{\nu}\gamma_{5}\right)
        \Big(\mu({\bf k'},h')\overline{\mu}({\bf k'},h')\Big) 
  \right] \;,
\end{equation}
and a discussion of the hadronic tensor $W^{\mu\nu}$ is
postponed until the next section. 

We conclude this section with the evaluation of the leptonic tensor. 
To do so, both matrices $\nu\overline{\nu}$ and $\mu\overline{\mu}$ 
in Eq.~(\ref{eq_12}) are first expressed in terms of Dirac matrices. 
For the case of the massive muon we obtain
\begin{equation}
\label{eq_15}
 \mu({\bf k'},h')\overline{\mu}({\bf k'},h')=
 \frac{\left(\rlap/k'+m\right)}{2E_{k'}}
 \left[\frac{1}{2}(1+h'\gamma^{5}\rlap/s)\right] \;,
\end{equation}
where the four-component spin vector is given by
\begin{equation}
\label{eq_16}
 s^{\mu} \equiv s^{\mu}({\bf k}')=
 \frac{1}{m}\Big(k',E_{k'}\,\widehat{\bf k}'\Big) \;.
\end{equation}
The corresponding expression for the massless left-handed neutrino 
may be obtained from the above equations by setting the helicity to 
$h\!=\!-1$ and by taking the massless ($m\!\rightarrow\!0$) limit.
Note that in the massless limit $m s^{\mu}\rightarrow k^{\prime\mu}$. 
Thus we obtain,
\begin{equation}
\label{eq_14}
 \nu({\bf k})\overline{\nu}({\bf k})=
 \frac{\rlap/k}{2 E_{k}}
 \left[\frac{1}{2}(1 + \gamma^{5})\right] \;
\end{equation}
Finally, by substituting the above expressions into the leptonic 
tensor of Eq.~(\ref{eq_12}), which in turn we separate into 
($\mu\leftrightarrow\nu$) symmetric and antisymmetric parts, 
\begin{equation}
\label{eq_17}
 {\ell^{\,\mu\nu}} \equiv {\ell}^{\,\mu\nu}_{S} 
                   +      {\ell}^{\,\mu\nu}_{A}\;,
\end{equation}
we obtain
\begin{subequations}
 \begin{equation}
  \label{eq_18}
  {\ell}^{\,\mu\nu}_{S}=
  \frac{2}{kE_{k'}}
   \Big(k^{\mu}K^{\prime\nu}+K^{\prime\mu}k^{\nu}-
        g^{\mu\nu }k \cdot K'\Big) \;,
 \end{equation}
 \begin{equation}
  \label{eq_19}
  {\ell}^{\,\mu\nu}_{A}=-
  \frac{2i}{kE_{k'}}
  \varepsilon^{\mu\nu\alpha\beta}k_{\alpha}K'_{\beta}\;.
 \end{equation}
 \label{eq_18_19}
\end{subequations}
Note that in the above expressions the following four-vector has 
been introduced:
\begin{equation}
 \label{KPrimeDef}
  K' \equiv \frac{1}{2}(k'-h'ms)
 \mathop{\longrightarrow}_{m=0} k'\delta_{h',-1}\;,
\end{equation}
where the last expression denotes the massless limit.  Hence,
in the $m\!\rightarrow\!0$ limit the leptonic tensor vanishes for
positive-helicity ($h'\!=\!+1$) but for negative-helicity
($h'\!=\!-1$) goes over to Eq.~(17) of
Ref.~\cite{VanderVentel_PRC69_2004}. Finally, note that the following
convention was adopted~\cite{Yndurain99}:
\begin{equation}
 {\rm Tr}\left(\gamma^{5}\gamma^{\mu}\gamma^{\nu}
               \gamma^{\alpha}\gamma^{\beta}\right)
           =4i\,\varepsilon^{\mu\nu\alpha\beta}\;,
         \quad (\varepsilon^{0123}=-1, \;
                \varepsilon_{0123}=+1) \;.
 \label{epschoice}
\end{equation}
We close this section with a comment on the conservation (or rather
lack-thereof) of the leptonic tensor. While the antisymmetric
component satisfies
\begin{equation}
\label{eq_20}
 q_{\mu}{\ell}^{\,\mu\nu}_{A}={\ell}^{\,\mu\nu}_{A}q_{\nu}=0\;,
\end{equation}
due to the antisymmetric property of the Levi-Civita tensor, this 
is no longer true for the symmetric part due to the finite muon 
mass. That is,
\begin{equation}
\label{eq_21}
 q_{\mu}{\ell}^{\,\mu\nu}_{S} \neq 0
 \quad {\rm and} \quad
 {\ell}^{\,\mu\nu}_{S}q_{\nu} \neq 0\;.
\end{equation}

\subsection{Differential cross section in terms of nuclear structure functions}
\label{section_xsection_structure}

In Eq.~(\ref{eq_11}) of the previous section it was shown that the 
differential cross section for the CC reaction may be written as 
a contraction of the leptonic tensor with the hadronic tensor, where 
the latter is defined in terms of the expectation value of the weak 
nuclear operator [see Eq.~(\ref{eq_5})]. That is,
\begin{equation}
 W^{\mu\nu} = \left[\langle p',s' \, ; \Psi_{f}(P') 
                   \Big| \hat{J}^{\mu}(q) \Big|  
              \Psi_{i}(P)\rangle \right]
              \left[\langle p',s' \, ; \Psi_{f}(P') 
                   \Big| \hat{J}^{\nu}(q) \Big|  
              \Psi_{i}(P)\rangle \right]^{*}
            \equiv W_{S}^{\mu\nu} + W_{A}^{\mu\nu}\;.
 \label{Wmunu}
\end{equation}
Although the general form of the hadronic tensor was introduced and
discussed in detail in Ref.~\cite{VanderVentel_PRC69_2004}, some of 
its most salient features are underscored here for completeness. For
the case of unpolarized proton emission, the hadronic tensor may be 
written in terms of thirteen independent structure functions,
\begin{subequations}
\begin{eqnarray}
 W_{S}^{\mu\nu} & = &
 W_{1} g^{\mu \nu}+ W_{2} q^{\mu}q^{\nu}+W_{3} P^{\mu}P^{\nu}+
 W_{4}\, p^{\prime\mu}p^{\prime\nu} \nonumber \\
                & + &
 W_{5} (q^{\mu}P^{\nu}\!+\!P^{\mu}q^{\nu})+
 W_{6} (q^{\mu}p^{\prime\nu}\!+p^{\prime\mu}q^{\nu})+
 W_{7} (P^{\mu}p^{\prime\nu}\!+p^{\prime\mu}P^{\nu})\;,
 \label{eq_hadron_tensora} \\
 W_{A}^{\mu\nu} & = &
 W_{8}  (q^{\mu}P^{\nu}\!-\!P^{\mu}q^{\nu}) +
 W_{9}  (q^{\mu}p^{\prime\nu}\!-p^{\prime\mu}q^{\nu})+
 W_{10} (P^{\mu}p^{\prime\nu}\!-p^{\prime\mu}P^{\nu}) \nonumber \\
                & + &
 W_{11} \varepsilon^{\mu\nu\alpha\beta}q_{\alpha}P_{\beta} +
 W_{12} \varepsilon^{\mu\nu\alpha\beta}q_{\alpha}p'_{\beta} +
 W_{13} \varepsilon^{\mu\nu\alpha\beta}P_{\alpha}p^{\prime}_{\beta}\;.
\label{eq_hadron_tensorb}
\end{eqnarray}
\end{subequations}
Note that all structure functions are functions of the four
Lorentz-invariant quantities, $q^{\mu}q_{\mu}\!\equiv\!-Q^{2}$,
$q\cdot P$, $q\cdot p^{\prime}$, and $P\cdot p^{\prime}$. Details 
on the contraction between the leptonic and hadronic tensors
\begin{eqnarray}
\label{eq_22}
 {\ell}_{\mu \nu} \, W^{\mu \nu}
 & = & 
 {\ell}_{\mu\nu}^{S}W^{\mu \nu}_{S} + 
 {\ell}_{\mu\nu}^{A}W^{\mu \nu}_{A} \;,
\end{eqnarray}
have been reserved to Appendix~\ref{section_appendix1}. Yet we note
that the charged-current reaction is now sensitive to the three
structure functions $W_{2}$, $W_{5}$ and $W_{6}$. This is in contrast
to the neutral-current reaction (see Eq.~(22a) of
Ref.~\cite{VanderVentel_PRC69_2004}); the origin of this difference is
the non-conservation of the symmetric part of the leptonic tensor due
to the finite muon mass [see Eq.~(\ref{eq_21})]. However, as the
antisymmetric part of the leptonic tensor is manifestly conserved,
both NC and CC processes are insensitive to the $W_{8}$ and $W_{9}$
structure functions.

This concludes the model-independent description of charged-current
neutrino-nucleus scattering. In summary, the cross section may be 
parametrized in terms of eleven nuclear-structure functions. In 
principle, they could be determined by a ``super'' Rosenbluth 
separation. In practice, however, this is not possible so we resort 
to a relativistic mean-field model to obtain explicit expressions for 
these quantities. This will be done in the next section.

\subsection{Model-dependent evaluation of the cross section}
\label{section_xsection_model}

In the previous section a model-independent formalism was presented for
charged-current neutrino-nucleus scattering. Specifically, the cross 
section was written in terms of a set of nuclear structure functions 
that parametrize our ``ignorance'' about the strong-interactions physics 
at the hadronic vertex. However, to proceed any further a number of 
approximations need to be made in order to obtain a numerically tractable 
problem.

The first ``no-recoil'' approximation, detailed in Eqs.~(23)--(26) 
of Ref.~\cite{VanderVentel_PRC69_2004}, is purely kinematical and 
yields the following expression for the angle-integrated differential 
cross section:
\begin{equation}
 \label{eq_29}
 \frac{d\sigma(h')}{dE_{p'}} = 
  \frac{G_{F}^{\,2}\cos^{2}\theta_{C}}{2(2\pi)^{4}}
  \,k^{\prime}E_{k^{\prime}}p^{\prime}E_{p^{\prime}}
  \int_{0}^{\pi}\sin\alpha\,d\alpha  
  \int_{0}^{\pi}\sin\theta\,d\theta  
  \int_{0}^{2\pi}d\phi\,\Big({\ell}_{\mu\nu}W^{\mu \nu}\Big) \;.
\end{equation}
Here $\alpha$ is the polar angle defining the direction of the
outgoing proton having momentum $p'\!\equiv\!|{\bf p}'|$ and energy
$E_{p^{\prime}}\!=\!\sqrt{p^{\prime 2}+M^{2}}$. Similarly, $\theta$
and $\phi$ define the polar and azimuthal angles of the outgoing muon
with momentum $k'\!\equiv\!|{\bf k}'|$ and energy
$E_{k^{\prime}}\!=\!\sqrt{k^{\prime2}+m^{2}}$. (For further details we
refer the reader to Fig.~2 of Ref.~\cite{VanderVentel_PRC69_2004}).
Finally, to compare the present charged-current calculation to the
neutral-current one, we have integrated over the kinematic variables
of the outgoing lepton.

The second approximation concerns the evaluation of the nuclear matrix 
element
\begin{equation}
\label{eq_30}
 J^{\mu} = \langle p',s' \, ; \Psi_{f}(P') 
           \Big| \hat{J}^{\mu}(q) \Big|  
           \Psi_{i}(P)\rangle \;.
\end{equation}
First, two- and many-body components of the current operator are 
neglected by assuming that the $W$-boson only couples to a single 
bound neutron. Second, two- and many-body rescattering processes 
are neglected by assuming that the detected proton is associated 
with the specific bound neutron to which the $W$-boson had coupled
to. Further, as we are confident that distortion effects largely
factor out from the ratio of cross
sections~\cite{Meucci_2004,Mar05_xxx0505008}, final-state interactions
between the outgoing proton and the residual nucleus will be
neglected. Finally, the impulse approximation is invoked by assuming
that the weak charged-current operator for a nucleon in the nuclear
medium retains its free-space form. That is,
\begin{equation}
\label{eq_31}
 \hat{J}_{\mu} \equiv \hat{J}_{\mu}^{\rm CC}
               -      \hat{J}_{\mu 5}^{\rm CC}
               =  F_{1}(Q^{2})\gamma_{\mu} 
               + iF_{2}(Q^{2})\sigma_{\mu\nu}\frac{q^{\nu}}{2M}  
               -  G_{A}(Q^{2})\gamma_{\mu}\gamma_{5}\;.
\end{equation}
Here $M$ is the nucleon mass and $F_{1}$, $F_{2}$, and $G_{A}$ are
Dirac, Pauli, and axial-vector nucleon form factors.  Note that the
pseudoscalar form factor has been neglected, since its contribution is
suppressed by the small lepton mass~\cite{Kim_PRC51_1995}.  A detailed
discussion of the weak charge current [Eq.~(\ref{eq_31})] has been
reserved to Appendix~\ref{section_appendix2}.  As we have assumed that
the ratio of cross sections given in Eqs.~(\ref{eq_1})
and~(\ref{eq_2}) are insensitive to final state interactions between
the outgoing proton and the residual nucleus, both initial (bound) and
final (free) nucleon propagators may be written in terms of Dirac
gamma matrices --- rendering the hadronic tensor analytical. Explicit
expressions for both propagators and for the analytic (albeit
model-dependent) hadron tensor are given in Eqs.~(34--38) of
Ref.~\cite{VanderVentel_PRC69_2004}.

\section{Results} 
\label{section_results}

\begin{figure}[ht]
\includegraphics[height=8cm,angle=0]{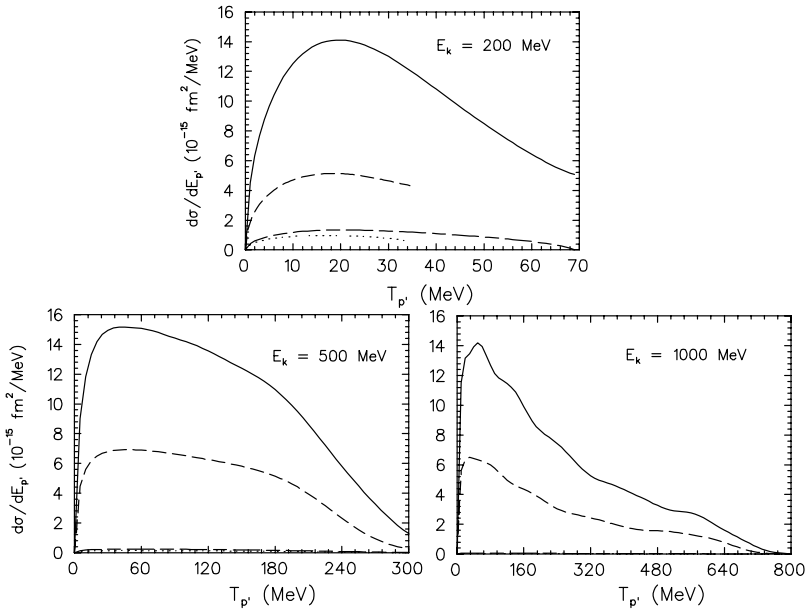}
\caption{Differential cross section $d \sigma/dE_{p'}$
[Eq.~(\ref{eq_29})] as a function of the outgoing proton laboratory
kinetic energy $T_{p'}$.  The solid and dashed lines denote the
contributions from the $1p^{3/2}$ and $1s^{1/2}$ neutron orbitals in
$^{12}$C to the negative-helicity $(h\!=\!-1)$ cross section.  The
long-dashed--short-dashed and dotted lines are the corresponding
contributions for positive helicity $(h\!=\!+1)$.  The three
considered incident neutrino energies are $E_{k}\!=\!200, 500$, and 
$1000$ MeV.}
\label{fig_2}
\end{figure}

In this section results are presented based on the formalism outlined
in Sec.~\ref{section_formalism} for charged-changing neutrino
scattering from ${}^{12}$C.  The angle-integrated differential cross
section [Eq.~(\ref{eq_29})] is shown in Fig.~\ref{fig_2} as a function
of the kinetic energy $T_{p'}$ of the outgoing proton in the
laboratory frame for three incident neutrino energies, namely,
$E_{k}\!=\!k\!=\!200, 500$ and $1000$ MeV. We display separately the
contribution to the cross section from the $1p^{3/2}$ (solid and
long-dashed--short-dashed lines) and $1s^{1/2}$ (dashed and dotted
lines) orbitals computed in a relativistic mean-field approximation
using the NL3 parameter set~\cite{Lalazissis_PRC55_97}. Note that
because of the finite muon mass, both negative ($h'\!=\!-1$) and
positive ($h'\!=\!+1$) helicity muons contribute to the cross section;
the two smallest contributions correspond to the positive-helicity
case.  As the energy of the incident neutrino increases, and
consequently also that of the outgoing muon, the positive-helicity
contribution (which scales as $m/E_{k'}$) becomes less and less
important until it ultimately disappears at large-enough energy. This
can already be observed at $E_{k}\!=\!500$ and $1000$ MeV.

For the elementary process $\nu + n \longrightarrow \mu^{-} + p$, the 
threshold laboratory energy of the incident neutrino is 
approximately 112 MeV.
An additional kinematical constraint that is strongly affected by
binding energy corrections follows from energy conservation. Using the
fact that $ E_{k'}\!\ge\!m$ we obtain
\begin{equation}
 \label{eq_35}
  E_{k}+(M+E_{B}) \ge m+(T_{p'}+M) 
  \implies
  T_{p'} \le E_{k}-E_{B}-m \;,
\end{equation}
where $E_B$ is the (positive) binding energy of the neutron. For
${}^{12}$C, the NL3 parameter set predicts 
$E_{B}(1s^{1/2})\!\approx\!53$ MeV and  
$E_{B}(1p^{3/2})\!\approx\!19$ MeV. For the particular case of a 
neutrino incident energy of $E_{k}\!=\!200$~MeV, the cross section
displays a sharp cut-off for the knockout of the $1s^{1/2}$ neutron at
an energy of $T_{p'}\!\approx\!40$~MeV. For higher incident neutrino
energies, the maximum allowed value for the kinetic energy of the
outgoing proton is already sufficiently large to allow the cross
section to fall off smoothly to (almost) zero. Our subsequent results
will only focus on incident neutrino energies of 500 and 1000 MeV, as
the ratio $R_{NC/CC}$ defined in Eq.~(\ref{eq_2}) will be measured by
the FINeSSE collaboration with neutrinos in that energy 
range~\cite{Brice_2004}. 

\begin{figure}[ht]
\includegraphics[height=8cm,angle=0]{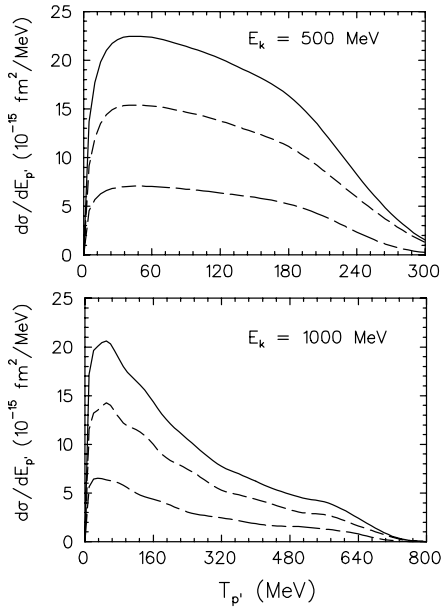}
\caption{Differential cross section $d \sigma/d E_{p'}$
[Eq.~(\ref{eq_29})] as a function of the outgoing proton laboratory
kinetic energy $T_{p'}$.  The target nucleus is $^{12}$C and the
incident neutrino energy is taken to be 500 and 1000 MeV.  The solid
line represents the cross section calculation summed over both muon
helicities ($h\!=\!\pm 1$) and over both bound-state ($1s^{1/2}$ and
$1p^{3/2}$) orbitals. The dashed and long-dashed--short-dashed lines
represent the individual contributions to the (unpolarized muon) cross
section from the $1p^{3/2}$ and $1s^{1/2}$ orbitals, respectively.}
\label{fig_3}
\end{figure}

The cross section that results from adding the contributions from both
muon helicities ($h\!=\!\pm 1$) and both neutron orbitals ($1s^{1/2}$
and $1p^{3/2}$) is depicted in Fig.~\ref{fig_3} by the solid line.
The dashed and long-dashed--short-dashed lines represent the
calculation where we have summed over the two helicity values for the
individual $1p^{3/2}$ and $1s^{1/2}$ orbitals, respectively. The
result for the full cross section (solid line) may be compared to
Fig.~8 of Ref.~\cite{Alberico_NPA623_97}. In the kinematical region 
in which they can be compared, there is good agreement in both the 
shape and magnitude of the cross sections.

\begin{figure}[ht]
\includegraphics[height=8cm,angle=0]{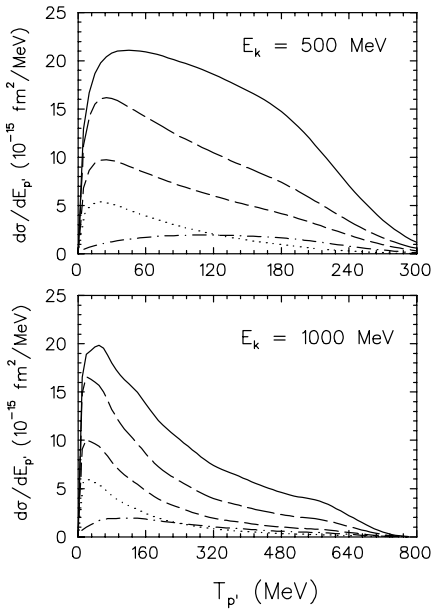}
\caption{Effect of the single-nucleon form factors on the differential
cross section [Eq.~(\ref{eq_29})] as a function of the laboratory
kinetic energy $T_{p'}$ of the outgoing proton.  The calculations
include a sum over the two ($1s^{1/2}$ and $1p^{3/2}$) neutron
orbitals in $^{12}$C and are for incident neutrino energies of 500 and
1000 MeV.  Explanation for the various lines is given in the text.}
\label{fig_4}
\end{figure}

Next we investigate in Fig.~\ref{fig_4} the contribution from the
single-nucleon form factors to the differential cross section for 
incident neutrino energies of $E_{k}\!=\!500$ and $1000$~MeV. As in 
the previous figures, the full result is displayed by the solid line. 
Next in importance is the long-dashed--short-dashed line obtained by
setting the weak Pauli form factor to zero ($F_{2}\!\equiv\!0$). The
last three lines are obtained from calculations using a single
non-zero form factor. That is, the dashed line is obtained from the
full calculation by setting $F_{1}\!=\!F_{2}\!=\!0$, the dotted line
by setting $G_{A}\!=\!F_{2}\!=\!0$, and the dashed-dotted line by
setting $G_{A}\!=\!F_{1}\!=\!0$.  This figure clearly illustrates the
relatively minor role played by the kinematically suppressed weak
Pauli form factor $F_{2}$.  Indeed, by itself, it yields a partial
cross section that both in magnitude and in shape shows little
resemblance to the full cross section. Clearly, the two dominant form
factors are the weak Dirac and the axial-vector form factors, with the
latter assuming the dominant role. Yet by itself, no single form
factor reproduces the full cross section indicating that {\it all}
interference terms, $F_{1}F_{2}$, $F_{1}G_{A}$, and $F_{2}G_{A}$ are
important for the charged-current process. Contrast this to the
neutral-current process where the Dirac form factor is
strongly suppressed by the weak mixing angle
($1\!-\!4\sin^{2}\theta_{W}\!\approx\!0.076$).

As mentioned earlier, systematic errors with the neutron detection
make the ratio of neutral- to charged-current reactions $R(NC/CC)$ a
more viable alternative than the proton-to-neutron neutral-current
ratio $R(p/n)$. Thus, we now compare in Fig.~\ref{fig_5} the cross
section for the charged-current reaction with that for the
neutral-current process: $\nu\!+\!X(Z,A) \longrightarrow
\nu\!+\!p\!+\!X(Z\!-\!1,A\!-\!1)$.  The solid line represents the full
charged-current cross section from ${}^{12}$C, where we have summed
over both bound-state orbitals and both muon helicities. The dashed
line represents the corresponding neutral-current cross section with
no strange-quark contribution to the spin of the proton ({\i.e.,
$g_{A}^{s}\!\equiv\!0$); the long-dashed--short-dashed line is the
same calculation with $g_{A}^{s}\!=\!-0.19$. Results are shown for
incident neutrino energies of 500~MeV (top graph) and 1000~MeV (bottom
graph).
\begin{figure}[ht]
\includegraphics[height=8cm,angle=0]{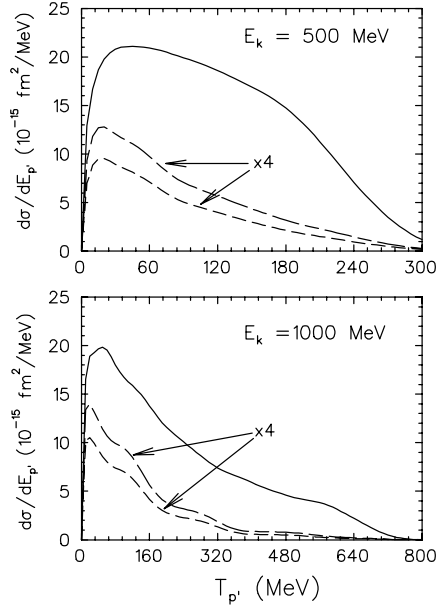}
\caption{Differential cross section $d \sigma/d E_{p'}$
[Eq.~(\ref{eq_29})] for neutrinos on $^{12}$C as a function of the
laboratory kinetic energy $T_{p'}$ of the outgoing proton. The solid
line represents the full charged-current neutrino-nucleus cross
section, while the dashed (long-dashed--short-dashed) line represents
the corresponding neutral-current neutrino-nucleus cross section using
$g_{A}^{s}\!=\!0$ ($g_{A}^{s}\!=\!-0.19$).  Results are shown for
incident neutrino energies of 500 and 1000 MeV. Note that for clarity
the neutral-current cross sections have been multiplied by a factor 
of 4.}
\label{fig_5}
\end{figure}
A comparison to Fig.~8 of Ref.~\cite{Alberico_NPA623_97} shows good
agreement in both the shape and magnitude of the cross sections.  The
axial-vector form factor plays a dominant role in the neutral-current
neutrino-proton reaction and makes this reaction particularly
sensitive to the strange-quark contribution to the spin of the
proton. Recall that the axial-vector form factor of a proton in the
neutral-current case is given by~\cite{VanderVentel_PRC69_2004}
\begin{equation}
\label{eq_36}
 \widetilde{G}_{A}(Q^{2})=\Big(g_{A}-g_{A}^{s}\Big)G^{A}_{D}(Q^{2})
            \mathop{\longrightarrow}_{Q^{2}\,=\,0} (1.26-g_{A}^{s})
            \mathop{\longrightarrow}_{g_{A}^{s}\,=\,-0.19} 1.45\;.
\end{equation}
Here $G^{A}_{D}(Q^{2})$ is the axial-vector form factor of the nucleon
(see Appendix~\ref{section_appendix2}) and a value of
$g_{A}^{s}\!=\!-0.19$ is assumed for the strange-quark contribution to
the spin of the nucleon~\cite{Horowitz_PRC48_93}; this value seems to
improve the agreement with the Brookhaven National Laboratory
experiment E734~\cite{Ahrens_PRD35_87}. Note that this negative value
of $g_{A}^{s}$ leads to an increase in the proton $\widetilde{G}_{A}$
by about 15\%.
\begin{figure}[ht]
\includegraphics[height=10cm,angle=0]{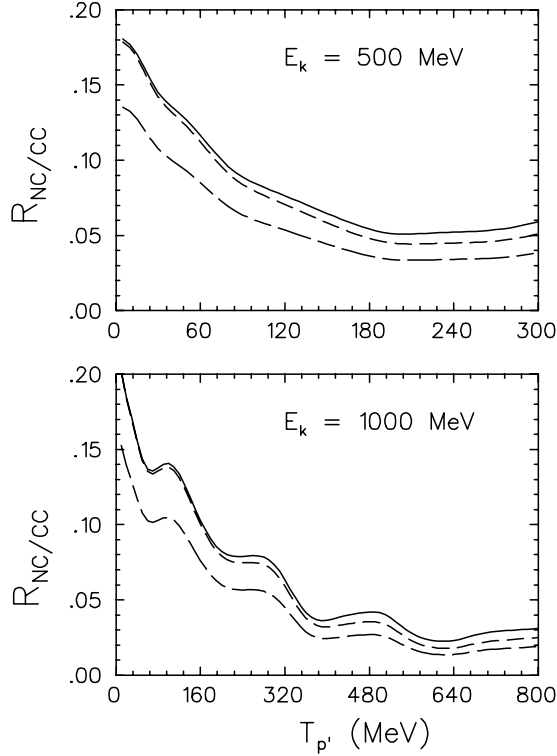}
\caption{Ratio of neutral- to charged-current neutrino-nucleus
($^{12}$C) cross sections as a function of the laboratory kinetic
energy of the outgoing proton $T_{p'}$. The solid and
long-dashed--short-dashed lines correspond to $g_{A}^{s}\!=\!-0.19$
and $g_{A}^{s}\!=\!0$, respectively.  The dashed line is obtained by
multiplying the $g_{A}^{s}\!=\!0$ result by a constant factor of 1.32
[see Eq.~(\ref{eq_38})].  The incident neutrino energies are 500 and
1000 MeV.}
\label{fig_6}
\end{figure}

We are now in a position to display results for the main observable of
this work: the ratio of neutral- to charged-current neutrino-nucleus
scattering cross sections $R(NC/CC)$ defined in Eq.~(\ref{eq_2}). For
notational simplicity let us denote the differential cross section
$d\sigma/dE$ by $\sigma$, where it is implied that we have summed over
the $1s^{1/2}$ and $1p^{3/2}$ orbitals of $^{12}$C as well as over the
two values of the helicity (when appropriate). Then, because of the
dominance of the axial-vector form factor we may write the NC cross
sections as
\begin{equation}
\label{eq_37}
 \frac{\sigma_{NC}(g_{A}^{s}\!\neq\!0)}{\sigma_{NC}(g_{A}^{s}\!=\!0)}
  \approx \left(1-\frac{g_{A}^{s}}{g_{A}}\right)^{2}\;.
\end{equation}
As the strange-quark contribution to the spin of the nucleon is assumed
to be isoscalar, the charged-current reaction is insensitive to it. Thus,
\begin{equation}
\label{eq_38}
 \frac{R(NC/CC;g_{A}^{s}\!\neq\!0)}
      {R(NC/CC;g_{A}^{s}\!=\!0)} \approx
  \left(1-\frac{g_{A}^{s}}{g_{A}}\right)^{2}
  \mathop{\longrightarrow}_{g_{A}^{s}\,=\,-0.19} 1.32 \;.
\end{equation}
That is, assuming the dominance of the axial-vector form factor in
neutral-current neutrino-proton scattering, a $\sim 30$\% enhancement
(for $g_{A}^{s}\!=\!-0.19$) in $R(NC/CC)$ is expected from a non-zero
strange-quark contribution to the spin of the nucleon. The $R(NC/CC)$
ratio is plotted in Fig. \ref{fig_6} as a function of the laboratory
kinetic energy $T_{p'}$ of the outgoing proton for incident neutrino
energies of 500 and 1000 MeV. The solid and long-dashed--short-dashed
lines correspond to non-zero and zero values of $g_{A}^{s}$,
respectively.  Further, the dashed line has been obtained by
multiplying the $g_{A}^{s}\!=\!0$ result by a constant enhancement
factor of 1.32.  The agreement between the solid and dashed lines
indicates that the simple estimate given in Eq.~(\ref{eq_38}) is
quantitatively correct --- especially at small $T_{p'}$ (or
equivalently small momentum transfer $q$) where the contribution from
the interference term $\widetilde{F}_{2}\widetilde{G}_{A}$ remains
small. While significant, the sensitivity to $g_{A}^{s}$ in
$R_{NC/CC}$ is about a factor of two less than in $R(p/n)$ where both
proton and neutron NC cross sections are sensitive to $g_{A}^{s}$.  We
trust that after working out the systematic uncertainties in the
neutron detection, this crucial experiment will also be performed.

We close this section with a brief comment on the mild oscillations
displayed by the neutral- to charged-current ratio $R_{NC/CC}$,
especially at 1000 MeV. Note that this structure is not unique to the
CC cross sections (see Fig.~\ref{fig_3}) but has already been observed
in the NC cross sections of Ref.~\cite{VanderVentel_PRC69_2004} (see
Fig.~9). As neither the momentum distribution of the bound nucleons
nor the nucleon form factors display any such structure, we attribute
this behavior to a kinematical effect originally pointed out in
Ref.~\cite{Horowitz_PRC48_93} (see Fig.~1) and later reproduced by us
in Figs.~5-6 of Ref.~\cite{VanderVentel_PRC69_2004}. The mild
oscillations in the single-differential cross section
$d\sigma/dT_{p'}$ is a {\it residual} effect associated with the
existence of a ``double-humped'' structure in the double-differential
cross section $d^{2}\sigma/dT_{p'}d( \cos \alpha)$ (here $\alpha$ is the
polar angle of the outgoing proton with kinetic energy $T_{p'}$).  In
turn, the emergence of the double-humped structure is a purely
kinematical effect that results from the inability of the reaction to
produce medium-energy nucleons.  That is, high-energy neutrinos are
able to produce low- or high-energy nucleons but not medium energy
ones. While the double-humped structure is a robust kinematical effect,
the integration over $\alpha$ introduces model dependences that smooth
out --- to a greater or lesser degree --- some of the structure
displayed by $d^{2}\sigma/dT_{p'}d( \cos \alpha)$.

\section{Summary}
\label{section_summary}

The distribution of mass, charge, and spin in the proton are among the
most fundamental properties in hadronic structure. In this context, a
topic that has received tremendous attention for over fifteen years is
the contribution of strange quarks to the structure of the proton.  In
this work we have focused on the strange-quark contribution to the
spin of the proton ($g_{A}^{s}$). Elastic neutrino-proton scattering
at low momentum transfer is particularly well suited for this study,
as the axial-vector form factor of the proton --- the observable that
encompasses the spin structure of the proton [see Eq.~(\ref{eq_36})]
--- dominates this reaction. Indeed, the two ``competing'' Dirac and
Pauli form factors are strongly suppressed, the former by the weak
mixing angle ($1\!-\!4\sin^{2}\theta_{W}\!\approx\!0.076$) and the
latter by the nucleon mass ($|{\bf q}|/2M$). Yet in an effort to
reduce systematic uncertainties related to the neutrino flux, a
``ratio-method'' has been proposed to extract $g_{A}^{s}$. Two ratios
are particularly useful in this regard: a) proton-to-neutron yields in
elastic neutrino scattering [Eq.~(\ref{eq_1})] and b) 
neutral-current-to-charged-current yields [Eq.~(\ref{eq_2})]. While
the former shows a larger sensitivity to $g_{A}^{s}$, the latter is
insensitive to systematic errors associated with neutron detection.
As neutrino experiments involve extremely low count rates, these 
reactions use targets that consists of a combination of free protons
and nucleons bound into nuclei. Thus, nuclear-structure effects must
be considered.

In the present work we have extended the formalism developed in
Ref.~\cite{VanderVentel_PRC69_2004} for neutral-current
neutrino-nucleus scattering to the charged-current reaction. In
particular, cross-section ratios have been computed within a
relativistic plane wave impulse approximation.  Benefiting from work
done by others~\cite{Meucci_2004,Mar05_xxx0505008}, we justify the
omission of final-state interactions by the suggestion that while
distortion effects change the overall magnitude of the cross section,
they do so without a substantial redistribution of strength.
Nuclear-structure effects --- which enter in our formalism exclusively
in terms of the momentum distribution of the bound nucleons computed
at the mean-field level --- were incorporated via the accurately
calibrated relativistic NL3 parameter set~\cite{Lalazissis_PRC55_97}.
The validity of the plane-wave approximation yields theoretical cross
sections that may be displayed in closed, semi-analytic form.
Although the structure of the weak hadronic current is the same for
the neutral- and charged-current reactions, a few differences
emerge. First, the finite muon mass results in muons produced with
both negative and positive helicity. Further, a finite muon mass
produces cross sections that display a sharp cut-off for low values of
the incident neutrino energy.  However, for the range of neutrino
energies of interest to the FINeSSE collaboration ($500$--$1000$ MeV)
the positive-helicity contribution becomes negligible. Further, while
the same three nucleon form factors enter the neutral- and
charged-current reactions, their quantitative impact differs
considerably.  For example, while the Dirac form factor for the
neutral-current process is strongly suppressed
($\widetilde{F}_{1}(Q^{2}\!=\!0)\!=\!0.076$) it is large for the
charged-current process ($F_{1}(Q^{2}\!=\!0)\!=\!1$). Hence, no single
form factor dominates the charged-changing reaction. More importantly,
as the strange-quark content of the nucleon is assumed to be
isoscalar, the purely isovector CC reaction is insensitive to the
strange-quark content of the nucleon. This renders the ratio
$R(NC/CC)$ less sensitive to strange quark effects (by about a factor
of 2) than the neutral-current ratio $R(p/n)$.  Still, for the value
of $g_{A}^{s}\!=\!-0.19$ adopted in this
work~\cite{Horowitz_PRC48_93}, a 30\% enhancemenent in $R(NC/CC)$ is
obtained relative to a calculation with $g_{A}^{s}\!=\!0$. We note
that our results for the charged-current cross sections were compared
to similar calculations done in Ref.~\cite{Alberico_NPA623_97} and
good agreement was found in both the shape and the magnitude of the
cross section.

In summary, the sensitivity of the ratio of neutral- to
charged-current cross sections to the strange-quark contribution to
the spin of the nucleon $g_{A}^{s}$ was investigated in a relativistic
plane wave impulse approximation. The enormous advantage of this
formalism is that our theoretical results may be displayed in closed,
semi-analytic form. The central motivation behind this work is the
proposed FINeSSE program that aims to measure $g_{A}^{s}$ with
unprecedented accuracy via the neutral- to charged-current ratio
$R(NC/CC)$.  By adopting a value of $g_{A}^{s}\!=\!-0.19$, an increase
in this ratio of approximately 30\% was found relative to the
$g_{A}^{s}\!=\!0$ result. While sensitive, this is less so than the
corresponding ratio of proton-to-neutron yields $R(p/n)$ in neutral
current neutrino-induced reactions. This measurement, however, has
been hindered by difficulties associated with neutron detection. We
trust that this difficulty may be overcome so that this crucial
program may get off the ground.

\begin{acknowledgments}
B.I.S.v.d.V gratefully acknowledges the financial support of the
University of Stellenbosch and the National Research Foundation of
South Africa. This material is based upon work supported by the
National Research Foundation under Grant number GUN 2048567
(B.I.S.v.d.V) and by the United States Department of Energy under
Grant number DE-FG05-92ER40750 (J.P.).
\end{acknowledgments}
\vfill\eject

\appendix
\section{Leptonic-hadronic contraction}
\label{section_appendix1}

In Sec.~\ref{section_xsection_lepton_hadron} it was shown that the
charged-current cross section could be expressed as the contraction 
of a leptonic tensor $\ell^{\mu\nu}$ [Eq.~(\ref{eq_12})] with a 
hadronic tensor $W^{\mu\nu}$ [Eq.~(\ref{Wmunu})] written in a 
model-independent way in terms of thirteen independent structure 
functions. In this appendix we carry out the contraction, which
we separate into symmetric and antisymmetric parts. That is,
\begin{equation}
\label{contraction}
 \ell^{\mu\nu}W_{\mu \nu} =
 {\ell}_{\mu\nu}^{S}W^{\mu \nu}_{S} +
 {\ell}_{\mu\nu}^{A}W^{\mu \nu}_{A} \;.
\end{equation}
Here the symmetric part is given by
\begin{eqnarray}
\nonumber
 \left(\frac{4}{kE_{k'}}\right)^{-1} 
 {\ell}_{\mu\nu}^{S}W^{\mu \nu}_{S} &=&
     \Big(-W_{1}(k\cdot K')
          +W_{2}\,{f}_{1}(q)
          +W_{3}\,{f}_{1}(P)
          +W_{4}\,{f}_{1}(p') 
          \\ \label{eq_23}
  && \hspace{0.4cm}
        +\,W_{5}\,{f}_{2}(P,q)
          +W_{6}\,{f}_{2}(q,p')
          +W_{7}\,{f}_{2}(P,p')\Big)\;,
\end{eqnarray}
while the antisymmetric part by
\begin{equation}
\label{eq_24}
 \left(\frac{4}{kE_{k'}}\right)^{-1} 
 {\ell}_{\mu\nu}^{A}W^{\mu \nu}_{A} = i
     \Big(W_{10}\,\varepsilon^{\mu\nu\alpha\beta}
          k_{\mu}K'_{\nu}P_{\alpha}p'^{\beta}
         +W_{11}\,f_{3}(q,P) 
         +W_{12}\,f_{3}(q,p') 
         +W_{13}\,f_{3}(P,p')\Big) \;.
\end{equation}
Note that the following four-vector has been defined:
\begin{equation}
 \label{KPrime}
 K' \equiv \frac{1}{2}(k'-h'ms)
 \mathop{\longrightarrow}_{m=0} k'\delta_{h',-1}\;.
\end{equation}
Further, for simplicity the following three functions have been 
introduced:
\begin{subequations}
\begin{eqnarray}
 && f_{1}(x) = 2(k\cdot x)(K'\cdot x) 
             - x^{2}(k\cdot K') \;, \\
 && f_{2}(x,y) = (k\cdot x)(K'\cdot y)
               + (k\cdot y)(K'\cdot x)
               - (x\cdot y)(k\cdot K') \;,\\
 && f_{3}(x,y) = (k\cdot y)(K'\cdot x)
               - (k\cdot x)(K'\cdot y) \;.
\label{ffunctions}
\end{eqnarray}
\end{subequations}

From Eq.~(\ref{eq_23}) we see that the three structure functions
$W_{2}$, $W_{5}$ and $W_{6}$ do contribute to charged-current
neutrino-nucleus scattering, in contrast to the neutral-current case
(see Eq.~(22a) of Ref.~\cite{VanderVentel_PRC69_2004}). This is due to
the lack of conservation of the symmetric part of the leptonic tensor
as a result of the finite muon mass [see Eq.~(\ref{eq_18})].  Note,
however, that in the massless limit
${f}_{1}(q)\!=\!{f}_{2}(P,q)\!=\!{f}_{2}(q,p')\!=\!0$, as required.
Finally, due to the form of Eq.~(\ref{eq_19}), the charged-current
process remains insensitive to the two structure functions $W_{8}$ and
$W_{9}$.

\section{Single nucleon form factors}
\label{section_appendix2}

In Sec.~\ref{section_xsection_model} it was shown that in
the impulse approximation, the single nucleon current probed in 
the charge-changing reaction may be written in the following 
standard form:
\begin{equation}
\label{app2_1}
 \hat{J}_{\mu} \equiv \hat{J}_{\mu}^{\rm CC}
               -      \hat{J}_{\mu 5}^{\rm CC}
               =  F_{1}(Q^{2})\gamma_{\mu} 
               + iF_{2}(Q^{2})\sigma_{\mu\nu}\frac{q^{\nu}}{2M}  
               -  G_{A}(Q^{2})\gamma_{\mu}\gamma_{5}\;,
\end{equation}
where $F_{1}$, $F_{2}$, and $G_{A}$ are the Dirac, Pauli, and
axial-vector form factors, respectively and the pseudoscalar form
factor has been neglected. To understand the structure of the vector
form factors ($F_{1}$ and $F_{2}$) we invoke the conservation of the
vector current (CVC) hypothesis. To start, one parametrizes the
nucleon matrix elements of the isovector electromagnetic current
in the following standard form:
\begin{eqnarray}
\label{app2_2}
 &&
 \langle N({\bf p}',s',t')
  |\hat{J}_{\mu}^{\,\rm EM}(T\!=\!1)|
 N({\bf p},s,t)\rangle =
 \langle N({\bf p}',s',t')
  |\overline{q}\gamma_{\mu}\frac{\tau_{3}}{2}q| 
 N({\bf p},s,t)\rangle \nonumber \\ && \hspace{0.75in} 
 \overline{U}({\bf p}',s')
   \Big[F_{1}^{(1)}(Q^{2})\gamma_{\mu} 
     + iF_{2}^{(1)}(Q^{2})\sigma_{\mu\nu}\frac{q^{\nu}}{2M} 
   \Big]U({\bf p},s)\,\Big(\tau_{3}\Big)_{tt'}\;,
\end{eqnarray}
where $\overline{q}\!=\!(\overline{u},\overline{d})$ is an isospin
doublet of quark fields, and $F_{1}^{(1)}$ and $F_{2}^{(1)}$ are the
{\it isovector} Dirac and Pauli form factors of the nucleon,
respectively. In turn, these are given in terms of proton and neutron
electromagnetic form factors as follows:
\begin{equation}
 \label{app2_3}
  F_{i}^{(1)}(Q^{2}) = \frac{1}{2}
  \left(F_{i}^{(p)}(Q^{2})-F_{i}^{(n)}(Q^{2})\right)\;, \quad (i=1,2)\;.
\end{equation}
The CVC hypothesis is a powerful relation that assumes that the vector
part of the weak charge-changing current may be directly obtained from 
the isovector component of the electromagnetic current. That is,
\begin{equation}
 \label{app2_4}
  \hat{J}_{\mu}^{\,\rm EM}(T\!=\!1) = \hat{V}_{\mu}^{(3)} =
  \overline{q}\gamma_{\mu}\frac{\tau_{3}}{2}q \;, \quad
  \hat{J}_{\mu}^{\,\rm CC}(\pm) = 
   \hat{V}_{\mu}^{(1)} \pm i\,\hat{V}_{\mu}^{(2)}=
   \overline{q}\gamma_{\mu}\left(\frac{\tau_{1}\pm 
                                    i\,\tau_{2}}{2}\right)q\;.
\end{equation}
Thus, a determination of the electromagnetic form factors of the 
nucleon --- which has been done experimentally --- fixes the vector 
part of the charge-changing currents to:
\begin{equation}
\label{app2_5}
 \langle N({\bf p}',s',t')
  |\hat{J}_{\mu}^{\,\rm CC}(\pm)|
 N({\bf p},s,t)\rangle =
 \overline{U}({\bf p}',s')
   \Big[F_{1}^{(1)}(Q^{2})\gamma_{\mu} 
     + iF_{2}^{(1)}(Q^{2})\sigma_{\mu\nu}\frac{q^{\nu}}{2M} 
   \Big]U({\bf p},s)\,\Big(2\tau_{\pm}\Big)_{tt'}\;.
\end{equation}
In this way the vector form factors of Eq.~(\ref{app2_1}) are then 
simply given by
\begin{equation}
 \label{app2_6}
   F_{i}(Q^{2})=2F_{i}^{(1)}(Q^{2})
  =F_{i}^{(p)}(Q^{2})-F_{i}^{(n)}(Q^{2})\;, \quad (i=1,2)\;.
\end{equation}
Paraphrasing Ref.~\cite{Walecka95}: {\it CVC implies that the vector
part of the single-nucleon matrix element of the charge-changing weak 
current, whatever the detailed dynamic structure of the nucleon, can
be obtained from elastic electron scattering through the electromagnetic
interaction!}

A similar procedure may be followed to determine the axial-vector form 
factor $G_{A}$ in terms of the isovector axial-vector current. That
is,
\begin{equation}
 \label{app2_7}
  \hat{J}_{\mu5}^{\,\rm CC}(\pm) = 
   \hat{A}_{\mu}^{(1)} \pm i\,\hat{A}_{\mu}^{(2)}=
   \overline{q}\gamma_{\mu}\gamma_{5}\left(\frac{\tau_{1}\pm 
                                    i\,\tau_{2}}{2}\right)q\;,
\end{equation}
so that
\begin{equation}
\label{app2_8}
 \langle N({\bf p}',s',t')
  |\hat{J}_{\mu5}^{\,\rm CC}(\pm)|
 N({\bf p},s,t)\rangle \equiv G_{A}(Q^{2})
 \overline{U}({\bf p}',s')
       \gamma_{\mu}\gamma_{5} 
 U({\bf p},s)\,\Big(\tau_{\pm}\Big)_{tt'}\;.
\end{equation}
As before, the above expression neglects the contribution from the
pseudoscalar form factor. 

We finish this section by parameterizing the various nucleon form
factors in terms of their known $Q^{2}\!=\!0$ values times form
factors of a dipole form. This is identical to the procedure employed 
in Appendix~A of Ref.~\cite{VanderVentel_PRC69_2004}) for the 
neutral-current reaction. We obtain
\begin{subequations}
\begin{eqnarray}
 && F^{(p)}_{1}(Q^{2})=
    \left(\frac{1+\tau(1+\lambda_{p})}{1+\tau}\right)
    G_{D}^{V}(Q^{2})\;, \quad
    F^{(p)}_{2}(Q^{2})=
    \left(\frac{\lambda_{p}}{1+\tau}\right)G_{D}^{V}(Q^{2})\;, \\
 && F^{(n)}_{1}(Q^{2})=
    \left(\frac{\lambda_{n}\tau(1-\eta)}{1+\tau}\right)
    G_{D}^{V}(Q^{2})\;, \quad
    F^{(n)}_{2}(Q^{2})=
    \left(\frac{\lambda_{n}(1+\tau\eta)}{1+\tau}\right)
    G_{D}^{V}(Q^{2})\;, \\
 && G_{A}(Q^{2})=g_{A}G_{D}^{A}(Q^{2})\;, 
\end{eqnarray}
\end{subequations}
where a dipole form factor of the following form is assumed:
\begin{subequations}
\begin{eqnarray}
  && G_{D}^{V}(Q^{2})=(1+Q^{2}/M_{V}^{2})^{-2}=(1+4.97\tau)^{-2}\; \\
  && G_{D}^{A}(Q^{2})=(1+Q^{2}/M_{A}^{2})^{-2}=(1+3.31\tau)^{-2}\; \\
  && \eta=(1+5.6\ \tau)^{-1} \quad \tau=Q^{2}/(4M^{2})\;.
\end{eqnarray}
\end{subequations}
Finally, for reference we display the value of the various nucleon
form factors at $Q^{2}\!=\!0$
\begin{subequations}
\begin{eqnarray}
 && F_{1}^{(p)}(0)=1\;, \quad
    F_{1}^{(n)}(0)=0\;, \\
 && F_{2}^{(p)}(0)=\lambda_{p}=+1.79\;, \quad
    F_{2}^{(n)}(0)=\lambda_{n}=-1.91\;, \\
 && G_{A}(0)=g_{A}=+1.26\;.
\end{eqnarray}
\end{subequations}
\vfill\eject

\bibliography{CC_paper} 

\end{document}